\documentstyle[twoside,fleqn,espcrc2,psfig]{article}

\catcode`@=11
\def\@maketitle{
  \global\setbox\fm@box=\vbox\bgroup	
    \vskip -30mm \normalsize	 	
    \rightline{UCSD-PTH-98-23}
    \rightline{hep-ph/9806309} 
    \rightline{May 1998}
    \normalsize
    \vskip 20mm                   
    \raggedright                  
    \hyphenpenalty\@M             
    {\Large \@title \par}         
    \vskip\@bls                   
    {\normalsize                  
     \@author \par}               
    \vskip\@bls                   
    \@address                     
  \egroup
  \twocolumn[%
    \unvbox\fm@box                
    \vskip\@bls                   
    \unvbox\abstract@box          
    \vskip 2pc]}                  
\catcode`@=12

\begin{document}

\title{Chirality, D-branes, and Gauge/String Unification}

\author{Erich Poppitz \address{Department of Physics,
University of California at San Diego,
La Jolla, CA 92093-0319, USA}\thanks{\tt epoppitz@ucsd.edu}}

\begin{abstract}
We point out that models where gauge and charged matter fields live on the
world volume of D3 branes can provide a weak string coupling description
of  the gauge coupling unification in the supersymmetric standard model at
a scale of order $10^{16}$ GeV. In the D3 brane picture, the string and 
unification scales coincide, while the size of the extra dimensions is
 somewhat larger than the string length.
We then construct a quasirealistic ``three generation" model with D3 branes
at orbifold fixed points.  We point out that placing D3 branes at 
different orbifold fixed points provides a geometric description
 of the disconnected branches of moduli space and 
discuss the appropriate consistency conditions. 
These disconnected branches allow for patterns of gauge symmetry 
breaking inaccessible on the connected part of moduli space and 
can give rise to   ``hidden" and ``visible" sectors.  
Finally,  we analyze supersymmetry breaking on the brane world volume.
(Based on a talk 
presented at the XXIIIrd 
Rencontres de Moriond ``{\it Electroweak Interactions and Unified Theories},"
Les Arcs, Savoie, France, March 14-21, 1998, and work with
J. Lykken and S.P. Trivedi, to appear).
\end{abstract}
 
\maketitle

\newcommand{\nc}{\newcommand}
\nc{\beq}{\begin{equation}} \nc{\eeq}{\end{equation}}
\nc{\beqa}{\begin{eqnarray}} \nc{\eeqa}{\end{eqnarray}}
\nc{\lsim}{\begin{array}{c}\,\sim\vspace{-21pt}\\< \end{array}}
\nc{\gsim}{\begin{array}{c}\sim\vspace{-21pt}\\> \end{array}}
\newcommand{\drawsquare}[2]{\hbox{%
\rule{#2pt}{#1pt}\hskip-#2pt
\rule{#1pt}{#2pt}\hskip-#1pt
\rule[#1pt]{#1pt}{#2pt}}\rule[#1pt]{#2pt}{#2pt}\hskip-#2pt
\rule{#2pt}{#1pt}}
\newcommand{\Yfund}{\raisebox{-.5pt}{\drawsquare{6.5}{0.4}}}
\newcommand{\Ysymm}{\raisebox{-.5pt}{\drawsquare{6.5}{0.4}}\hskip-0.4pt%
        \raisebox{-.5pt}{\drawsquare{6.5}{0.4}}}
\newcommand{\Yasymm}{\raisebox{-3.5pt}{\drawsquare{6.5}{0.4}}\hskip-6.9pt%
        \raisebox{3pt}{\drawsquare{6.5}{0.4}}}
\newcommand{\Ythree}{\raisebox{-3.5pt}{\drawsquare{6.5}{0.4}}\hskip-6.9pt%
        \raisebox{3pt}{\drawsquare{6.5}{0.4}}\hskip-6.9pt
        \raisebox{9.5pt}{\drawsquare{6.5}{0.4}}}

\def\inbar{\,\vrule height1.5ex width.4pt depth0pt}
\font\cmss=cmss12 \font\cmsss=cmss12 at 7pt
\def\IZ{\relax\ifmmode\mathchoice
{\hbox{\cmss Z\kern-.4em Z}}{\hbox{\cmss Z\kern-.4em Z}}
{\lower.9pt\hbox{\cmsss Z\kern-.4em Z}}
{\lower1.2pt\hbox{\cmsss Z\kern-.4em Z}}\else{\cmss Z\kern-.4em
Z}\fi}
\def\IB{\relax{\rm I\kern-.18em B}}
\def\IC{{\relax\hbox{$\inbar\kern-.3em{\rm C}$}}}
\def\ID{\relax{\rm I\kern-.18em D}}
\def\IE{\relax{\rm I\kern-.18em E}}
\def\IF{\relax{\rm I\kern-.18em F}}
\def\IG{\relax\hbox{$\inbar\kern-.3em{\rm G}$}}
\def\IP{\relax{\rm I\kern-.18em P}}
\def\ZN{{\IZ}_N}
\def\ZM{{\IZ}_M}
\def\Ztwo{{\IZ}_2}
\def\Zthree{{\IZ}_3}

\section{Introduction and summary.}

In the past several years, we have witnessed a tremendous progress
of our understanding of string theory. The five string
theories, previously thought to be different, have been 
shown to be related by a complicated web of dualities \cite{sen}. 
A crucial role in this understanding is played
by the extended nonperturbative objects in string theory, known as
D-branes  \cite{tasibranes}. 
D$p$-branes are $p$ spatial dimensional 
hypersurfaces embedded in ten dimensional space time, whose
world volumes support Yang-Mills theories with varying amount 
of supersymmetry. Furthermore, constructions 
using various branes give a unified description of many different  gauge 
theories in various dimensions  \cite{branereview} and 
provide some fascinating connections between string and gauge theory
dynamics, which are still being unraveled.

Superstring theory gives rise to gravity, gauge, 
and matter (fermion) fields, and is thus is a 
natural candidate for a theory unifying the known forces.
Most  model-building efforts involving strings have, so far, 
 been in the framework of the heterotic string theory. 
Since  all string theories are related by dualities, one may ask whether the 
weakly coupled description of the low-energy physics could be in terms of one of 
the other, dual, string theories. It is therefore interesting to 
investigate the model-building possibilities in the other string theories.

Another  motivation (admittedly, relying on certain assumptions) 
to look into different possibilities is provided by  the fact that 
the weakly coupled heterotic string has difficulties 
accommodating the ``observed" unification
of gauge couplings \cite{mssmgut} 
in the (supersymmetric) standard model at a  
scale $10^{16}$ GeV (for a review  see \cite{dienes}).
Witten proposed \cite{wittenstrongcy}  to consider instead 
the strong coupling dual
of the $E_8 \times E_8$ string---the eleven-dimensional $M$-theory 
compactified on a segment ($S^1/Z_2$) times a six-dimensional 
compact manifold (this ``$M$-theory phenomenology" 
has been the subject of some interest 
lately \cite{Mpheno}). Ref.~\cite{wittenstrongcy} 
also mentioned the possibility that Type I 
theories could  give a weak coupling 
description of the  gauge unification as well. 
Here we will elaborate somewhat on this issue and will point 
out some, hopefully, generic features of Type I ``realistic" models.

We will concentrate on models, where the gauge and matter fields
of the standard model live on the world volume of
three spatial dimensional extended objects 
(D3-branes) embedded in ten dimensional space time.
Since the closed strings (gravity) can also propagate in the bulk, 
in order to avoid violations of Newton's law at large distances the
six dimensions transverse to the D3 branes have to be compactified
 (a discussion on experimental limits on the 
sizes of the extra dimensions is given in ref.~\cite{ewstrings}). For
 recent related work on the subject, see refs.~\cite{dienes2}, \cite{kaku}.

This paper is structured as follows.
In Section 2, we begin with a discussion of gauge unification in models
with D3 branes. We will see that the ``observed" unification of couplings
in the supersymmetric extensions of the standard model can be accommodated
in models with D3 branes. The GUT and string scale are identified in these models,
while the size of the extra dimensions is somewhat larger than the inverse GUT
scale.  Then, in Section 3, we consider a simple compactification of 
Type I theory with D3 branes, considered  in \cite{sagnotti96} and more 
recently in \cite{tye}. 
We point out that, in addition to the connected part of moduli space, 
considered in \cite{sagnotti96}, this model exhibits disconnected 
branches of moduli space. These branches are best visualized in terms
of placing a set of D3 branes at orientifold planes away from the 
origin and are like the 
the disconnected vacua 
 discussed in \cite{wittentori}, \cite{rosly}. 
We also briefly discuss the consistency conditions \cite{SSW} 
 that these branches have to obey.
Along one of these branches, we obtain an $SU(5)$ model with three generations
of $\bf{\bar 5}$ and $\bf 10$ matter fields. 
The model is not realistic---there are no Higgs fields to break the either the 
$SU(5)$ or the standard model gauge groups, and there are baryon and lepton 
number violating Yukawa couplings. However, we view the ease with which this 
``quasirealistic" matter  content can be obtained in Type I 
compactifications as encouraging further study (for recent
work, see \cite{kaku}).
Finally, in Section 4, we analyze in some detail the infrared dynamics of the 
$SU(5)$ three generation model. We show that for fixed finite values of the 
dilaton and the orbifold blow-up parameter the model dynamically 
breaks supersymmetry.

\section{Gauge-string unification on D3 branes.}

We will show here that theories with gauge and matter
fields living on the world volume of
 D3 branes can, at weak string coupling,  accommodate the ``observed" 
unification of couplings of the (supersymmetric) standard model
 at a scale of order $10^{16}$ GeV.
We will see that in models
 with D3 branes, the string and ``GUT" scales are 
naturally identified.\footnote{Hereafter, 
by the ``grand unification" scale, $M_{GUT}$, 
we mean the scale where the gauge couplings become equal to one another;
one does not have to assume that there is a grand-unified gauge group.}
We will see that in the D3 brane picture,  
the extra dimensions are larger than the string scale 
and open up (in  energy units) somewhat below the string scale. They do 
not, however, affect the running of the gauge couplings, since the gauge 
and charged matter   fields are constrained to live on the world volume of the
D3 branes. At the string scale, new charged states appear in the 
spectrum---the massive states of the open  strings. The 
winding open string states are somewhat heavier than the string scale in this 
picture.

To find the relation between the string scale, string coupling, the size of
the compact dimensions, and the Planck  scale, 
consider the closed string effective action in 10 dimensions  
(see, e.g.~\cite{tasibranes}):
\beq
\label{gravity10d}
S_{grav, 10d} = {1\over 2} {1\over 64 \pi^7 \alpha^{\prime 4} g_s^2} 
\int d^{10} x {\cal R} + \ldots ,
\eeq
where $2 \pi \alpha^\prime$ is the string tension, and $g_s$---the string
coupling.
Upon dimensional reduction to 4 dimensions on a six-torus of volume 
$(2 \pi R)^6$ this becomes:
\beqa
\label{gravity4d}
S_{grav, 4d} &=& {1\over 2} {( 2 \pi R )^6
\over 64 \pi^7 \alpha^{\prime 4} g_s^2} 
\int d^4 x {\cal R} \nonumber \\
&=& {1\over 2} {M_{Pl}^2 \over 8 \pi} \int d^4 x {\cal
R} + \ldots ,
\eeqa
where $\sqrt{M_{Pl}^2/8 \pi} = 2.4 \times ~10^{18}$ GeV 
is the reduced Planck mass.
From eq.~(\ref{gravity4d}), we obtain the relation between the string scale, 
string coupling, and
the radius of the compact dimensions:
\beq
\label{gravityonD3}
{R^3 \over \sqrt{\pi} \alpha^{\prime 2} g_s}~=~ 2.4 \times 10^{18} ~{\rm GeV}~.
\eeq

The gauge and charged matter fields in models with Dp-branes are, on the
other hand, 
constrained to live on the brane world volume. Hence, to find the relation 
between the string coupling and scale, and the gauge coupling,
 we need to consider the world-volume theory, 
described by the Born-Infeld 
action \cite{tasibranes}:

$S_{BI} = $
$$
-{ 1 \over
  ( 2 \pi )^p \alpha^{\prime { p + 1 \over 2} } g_s } 
\int d^{p + 1} \sigma \sqrt{{\rm det}
 \left( G_{ab} +
 2 \pi \alpha^\prime~F_{a b} \right)} \
$$
\beqa
\label{borninfeld}
&=&
-{ ( 2 \pi \alpha^\prime)^2 \over
  ( 2 \pi )^p   \alpha^{\prime {p + 1 \over 2}} 
g_s } {1 \over 4} \int d^{p + 1} \sigma F_{ab} F^{ab} 
+ \ldots \nonumber \\ &=&
- {1 \over 4  g_{YM, p}^2} \int d^{p + 1} \sigma F_{ab} F^{ab} + \ldots.
\eeqa
Here $G_{ab}$ is the induced metric on the brane world volume, $F_{ab}$ is
the  gauge field strength, and we have kept only the terms describing the
Yang-Mills fields.
From eq.~(\ref{borninfeld}), we conclude that, on a Dp brane, 
the Yang-Mills gauge coupling is related to the string coupling by
\beq
\label{gaugeonDp}
g_{YM,p}^2 ~=~ g_s~(2 \pi)^{p -2}~ (\alpha^\prime)^{p -3 \over 2}~.
\eeq
This relation is valid at the string scale, $(\alpha^\prime)^{-1/2}$,
which is the cutoff scale of 
the effective world-volume theory.\footnote{We note that for nonabelian
fields, the relation (\ref{gaugeonDp}) can be modified by a normalization 
constant;
 it can be obtained by T-duality from the Type I
10-dimensional effective action, as in \cite{sen} (see also \cite{us}).}

From now on we consider D3 branes. 
The gauge coupling on 
the D3 brane at the string scale, eq.~(\ref{gaugeonDp}),  is:
\beq
\label{gaugecoupling}
\alpha_{YM} ~=~{ g^2_{YM}\over 4 \pi} ~=~{ g_s\over 2}~. 
\eeq
We  take the values of the unification scale $M_{GUT} \sim 10^{16}$
GeV and
the gauge coupling, $\alpha_{GUT} = 1/25$, for the
supersymmetric standard model \cite{mssmgut}.
From (\ref{gaugecoupling}) we thus obtain a value for 
the string coupling $g_s \sim .08$, which is well within the weak coupling 
regime.
On the other hand, taking 
$\alpha^\prime \sim M_{GUT}^{-2}$
from eq.~(\ref{gravityonD3}), 
we obtain $R M_{GUT} \simeq 3$. 
We see that identifying the string scale with the GUT scale in models with
D3 branes is natural---the size of the extra dimensions is larger than the 
string (and GUT) scale (which, as can be easily seen,
 is not the case in the T-dual 9-brane picture).
New states appear in the effective theory only at the string scale: the 
massive open and closed string excitations have mass of order 
$(\alpha^\prime)^{-1/2}$. The winding open string modes---the lightest
excitations
of open strings beginning and ending on the D3 branes, but winding around the
compact directions---on the
other hand, are somewhat 
heavier than the string scale, $M_{winding} \sim R M_{GUT}^2 \sim 3 M_{GUT}$.

\section{A ``three generation" chiral model on D3 branes.}

In this section, we present a simple ``three generation" model with D3 branes
on orbifold singularities (the orbifold projection will be useful, among other
things,  to obtain a chiral
gauge theory on the brane world-volume).
The model is, admittedly, not a realistic model, but it will serve 
the purpose of making
several generic points quite explicit. In addition to
providing a weak string coupling description of the ``observed" unification
of couplings,
the D3 brane picture has other potential benefits.

We will see that the D3 brane picture allows for a 
geometric description of the disconnected branches of
 moduli space of their world-volume theories.
These disconnected branches of moduli space correspond to placing D3 branes
at orbifold fixed points other than the origin (such fixed points only exist in 
compact orbifolds) and are like the ones discussed in  \cite{wittentori}, \cite{rosly}.
The disconnected branches of  moduli space 
can have multiple uses. We will see below that they exhibit  patterns of 
gauge symmetry breaking that are not possible on the 
connected part of moduli space. In addition, branes placed at  
different  orbifold
fixed points can serve as ``visible" and ``hidden" sectors; the latter can
be responsible for supersymmetry breaking. The lightest excitations of
strings stretching between branes at different fixed points transform as 
fundamentals under both the ``hidden" and ``visible" gauge groups; they
may be instrumental in  the communication of supersymmetry breaking.

As a simple example to illustrate the above points, 
we consider the $T^6/\Zthree$ 
orientifold in some detail. 
We take the D3 branes to stretch along $X^{1,2,3}$ and 
introduce  complex coordinates,
 $z_1 = X^4 + i X^5, z_2 = X^6 + i X^7, z_3 = X^8 + i X^9$,  in the 
transverse six-dimensional space.
The orientifold group  is given by
$
G= \{ 1, \alpha, \alpha^2 , \Omega  R  (-1)^{F_L} ,
\Omega R (-1)^{F_L} \alpha,
 \Omega R (-1)^{F_L} \alpha^2 \},$
where $\alpha = e^{2 \pi i/3}$, and the action on the transverse 
coordinates is 
\beq
\label{orbifold}
( z_1, ~ z_2, ~z_3 ) ~\rightarrow ( \alpha z_1, ~\alpha z_2, ~\alpha z_3 )~.
\eeq
$\Omega$ denotes world-sheet orientation reversal, and $R$ is a reflection 
$z_i \rightarrow - z_i, i = 1, 2, 3$ in the space transverse to the D3 branes.
$F_L$ is an operator that flips the sign of the left-moving Ramond states. 
In addition, the orientifold group acts on the Chan-Paton indices of the open
string states stretching between different D3 branes.
The action of the group elements,
$\Omega R (-1)^{F_L}$ and $\alpha$,  on the  Chan-Paton
factors $\lambda$ can be  represented by the matrices:  
\beq
\label{eomega}
\lambda \rightarrow \gamma_{\Omega R (-1)^{F_L}} ~ \lambda^T ~
\gamma_{\Omega R (-1)^{F_L}}^{-1}~,
\eeq
and 
\beq
\label{egamma} \lambda \rightarrow \gamma_\alpha ~ \lambda ~
\gamma_\alpha^{-1}.  
\eeq 
The matrices   $\gamma_\alpha $ and  $\gamma_{\Omega R (-1)^{F_L}}$ must  
furnish a
representation of the orientifold group. 
The matrices  $\gamma_{\Omega R (-1)^{F_L}}$,
representing the action of the $\IZ_2$ part of the orientifold group
should obey
\beq
\label{gammaomega}
\gamma_{\Omega R (-1)^{F_L}}\ =
\left( \gamma_{\Omega R (-1)^{F_L}} \right)^{T}~.
\eeq 
In the absence of the $\IZ_3$ orbifold projection,
 the $\Omega  R (-1)^{F_L}$ projection  would lead to 
 an $SO$ gauge group on the D3 brane world volume.

The  untwisted Ramond-Ramond 4-form 
charge conservation conditions require 
the presence of 32 D3 branes to cancel the orientifold charge.
On the other hand, the charge cancellation conditions for the
twisted RR fields 
result in the following requiremant on the matrices $\gamma_\alpha$:
\beq
\label{gammaalpha}
{\rm Tr} ~ \gamma_{\alpha} ~=~ -4 ~.
\eeq
This condition should be imposed on D3 branes placed at fixed points of
$\IZ_3$ (i.e. at the origin and the two other $\IZ_3$ fixed points in the
interior of $T^2$).

In Fig.~1 we show one of the three two-tori of this 
orbifold ($ T^6 = (T^2)^3$). On the
compact $T^2$ (as opposed to a noncompact $\IC$),
there is more than one fixed point of the orientifold $\IZ_6$
group. The origin, shown as a triangle on Fig.~1,
is the only $\IZ_6 = \IZ_2 \times
\IZ_3$ fixed point. In addition, there are three $\IZ_2$ fixed points
denoted by
squares, and two $\IZ_3$ fixed points, denoted by circles. Note that the 
three $\IZ_2$ fixed points are interchanged by the $\IZ_3$ action, while
the two $\Zthree$ fixed points are images under the $\Ztwo$ action.

\begin{figure}[ht]
\vspace*{13pt}
\centerline{\psfig{file=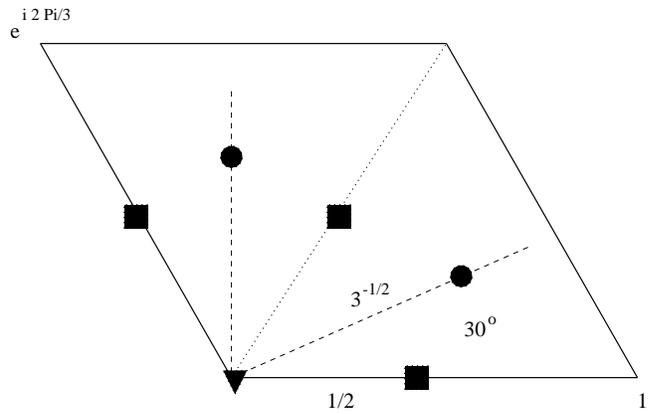}}
\vspace*{13pt}
\caption{ \small The $\IZ_6$ (triangle), $\IZ_3$ (circle), and 
$\IZ_2$ (square) fixed points on one of the transverse two-tori. The
``three-generation" $SU(5)$ model arises as the world volume theory
of the  11 D3 branes, placed
at the origin (the $\IZ_6$ fixed point). 
The remaining 21 D3 branes must be distributed among the other fixed points
or be at general positions, in a manner obeying some consistency conditions
(see text).
}
\end{figure}

Upon placing all 32 D3 branes at the origin,
we can easily recover \cite{us} the $SO(8)\times U(12)$ world volume
theory of \cite{sagnotti96}.
As these authors  pointed out, it is not possible to continuously break the
$SO(8) \times U(12)$ theory to an $SU(5)$ theory---in the D3 brane picture, as
we will see below, this corresponds to the fact that it is impossible to
smoothly move 21 of the 32 branes 
from the origin to the other fixed points; this is because 
the branes have to move in a $\IZ_6$ 
symmetric manner, i.e. in groups of 6.
We now note that the twisted RR charge cancellation condition 
can be satisfied with a smaller number, $n_0$, of D3 branes at the origin;
the general solution is:
$n_0 = 8 + 3p, p = 0,1,\ldots 8$.
Since there are two distinct values of $n_0$ mod 6, it appears that there
 can be two
disconnected branches of the moduli space. The minimal configuration with
$n_0$$=$$8$ is smoothly connected to the $n_0$$=$$32$ configuration.
The world volume theory has gauge group
$U(4)$ and matter content consisting of three antisymmetric
tensor representations ({\bf 6}'s) of $U(4)$, with no tree-level
superpotential. This can be obtained by a continuous breaking
of the $SO(8)\times U(12)$ world volume theory \cite{sagnotti96}.

We now ask whether it is possible to have  branches 
of the moduli space where an odd number of D3 branes is placed at the
$\IZ_2$ fixed points (i.e. the branch with 
$n_0 = 11$ mod $6$, discussed above). 
Such brane configurations would correspond to a 
disconnected branch of moduli space---as discussed above, an 
odd number of D3 branes can not be 
continuously removed from the origin. 
There are, however, some nonperturbative  consistency
conditions that such configurations have 
to obey \cite{SSW}, \cite{wittentori}. 
These are best 
elucidated in terms of turning on Wilson lines 
in the T-dual nine-brane theory; for a more detailed discussion, see \cite{us}. 
The nonperturbative consistency conditions have to do with the possibility 
to define spinors (which arise as nonperturbative states in the type-I theory)
on the corresponding gauge bundles \cite{SSW}.
The conditions require that the number of D3 branes 
in the $\IZ_2$ fixed points of each of 
the three two tori  be even. 
This, however, still allows  having 
an odd total number of branes removed from the origin. 
The simplest example involves  only two of the three
two-tori and
allows us to put 9 D3 branes in the $\IZ_2$ fixed points away 
from the origin is the following. 
We can place three (recall that they have
to be placed in a $\IZ_3$ symmetric manner)
of the 9 D3 branes in the
$\IZ_2$ fixed point in both the first and second torus. This configuration 
clearly does not obey the above consistency condition, 
since the number of D3 branes
at the $\IZ_2$ fixed points in each two-torus is odd. We can satisfy the 
condition by placing three additional branes at the $\IZ_2$ fixed points of
the first torus only, while keeping them at the origin of the second, and 
three additional branes at the $\IZ_2$ fixed points of the second torus, 
keeping them at the origin of the first. This configuration has 
  a total of 9 branes
at $\IZ_2$ fixed points and obeys the consistency condition. 
It is easy to see, that similarly, by displacing branes in all
the three two-tori, we can remove the desired 21 branes from the origin.

We can thus construct the 
 interesting ``three-generation" $SU(5)$ model: it  arises 
upon placing 11 D3 branes 
at the origin. Performing the orientifold projection it is easy to see that the
resulting $N=1$ theory has the following matter content:
\begin{equation}
\label{su5}
\begin{array}{c|c|c}
        &SU(5)&U(1)\\ \hline
A_{i = 1,2,3} & \Yasymm & 2 \\ \hline
 \bar{Q}_{i = 1,2,3} & \Yfund & - 1\\
\end{array} ~.
\end{equation}
The theory has a renormalizable tree-level superpotential given by
$W_{tree} = \epsilon^{ijk} A_i \bar{Q}_j \bar{Q}_k$.

We must distribute the remaining 21 D3 branes among 
the $\Ztwo$ and $\Zthree$
fixed points, and general points of the torus (accounting for 
the various consistency
conditions that need to be obeyed).
For the branes at the $\Zthree$ fixed points, the 
world volume theories are obtained by only imposing the $\Zthree$ projection, 
while the theories on the branes at the $\Ztwo$
fixed points
are obtained by imposing only the orientifold projection.
For branes at general points neither projection is required.
The world volume gauge theories from these three sets of branes have
different amounts of unbroken supersymmetries, since different
projections are imposed. The $SO(2k_2+1)$ gauge theory from branes at
the $\Ztwo$ fixed points has $N=4$ supersymmetry, as do the
$U(1)$ gauge theories from branes at general points. The
$[U(k_3)]^3$ gauge theory from branes at the $\Zthree$ fixed points
has $N=1$ supersymmetry. 

So far we have ignored the effects of open strings stretched between
branes at different fixed points. The lightest excitations of such
strings are massive states which transform as fundamental-antifundamental
under the respective world volume gauge groups. In the example we are
considering, there can be two world volume theories with $N=1$ supersymmetry.
If supersymmetry were dynamically broken in one of these theories,
supersymmetry breaking would be communicated to the other gauge
theory via the massive chiral multiplets just described (and, of course, by 
supergravity). 
A more precise investigation of this would probably
 involve considering details of the 
supersymmetry breaking dynamics and the stabilization of the  dilaton
\cite{dine};  we leave this for future work. 

\section{Supersymmetry breaking.}

In this section, we will consider briefly the infrared dynamics of
the $SU(5)$ model on the 11 D3 branes at the origin. We will show that,
for fixed value of the string coupling and 
for any finite value of the orbifold blow-up parameter
(i.e., the Fayet-Iliopoulos term of the anomalous $U(1)$), the ground
state of the world-volume theory dynamically breaks supersymmetry.
 The analysis, details of which can be found in \cite{us}, relies on
the fact that the dynamics of the $SU(5)$ three-generation model 
is known \cite{CSS}. The theory (\ref{su5}) is an s-confining theory, and the low
energy degrees of freedom are various composite meson and baryons:
\beqa
\label{mesons}
C ~=~& A \cdot \bar{Q} \cdot \bar{Q} &~\sim~ ( {\bf 3},~ {\bf \bar{3}} )~, 
\nonumber \\ 
B ~=~&A^5 &~\sim~ ({\bf 6}, ~{\bf 1}) ~,\\
 M ~=~&A^3 \cdot \bar{Q} &~ \sim~ ({\bf 8}, ~{\bf 3} ) ~, \nonumber 
\eeqa
where we have shown their transformation properties under the global
$SU(3)_A \times SU(3)_{\bar{Q}}$ symmetry. The confining 
superpotential is \cite{CSS}:
\beqa
\label{wconfining}
W &=&{ C_a^\alpha  B^{\beta \gamma}  M^{\delta a}_{\gamma} 
\epsilon_{\alpha \beta \delta} +  M^{\alpha a}_{\beta} M^{\beta b}_\gamma
M^{\gamma c}_\alpha \epsilon_{a b c} \over \Lambda^9} \nonumber \\
&+& 
\lambda  \delta^a_\alpha
C_a^\alpha ,
\eeqa
where $a, b, ... (\alpha, \beta, ...)$ 
denote indices under the $SU(3)_{{\bar{Q}} (A)}$ symmetry,
respectively, and the last term is the tree-level superpotential.
The tree-level
superpotential breaks the global symmetry to the diagonal $SU(3)_{diag}$
and lifts some of the classical flat directions (the moduli $B$ and some of the 
$M$ (\ref{mesons}) are not lifted). The superpotential coupling $\lambda$ in
(\ref{wconfining}) is proportional to the  value of the gauge 
coupling at the string scale (since the tree-level 
superpotential is the projection of the $N=4$ superpotential).

One can now show \cite{us} that (ignoring first 
the anomalous $U(1)$)
the theory (\ref{su5}) with superpotential (\ref{wconfining})
 has only a runaway supersymmetric solution (where the baryon $B \rightarrow
\infty$).
This suffices to show that the theory with the anomalous $U(1)$ 
breaks supersymmetry. 
To see this, note that the charges of all mesons and 
baryons under the $U(1)$ are positive.
The D-term of the anomalous $U(1)$, therefore, generates a
potential for the mesons and baryons that increases as
their expectation value increases. Depending on the sign and value of the
Fayet-Iliopolous term, this D-term potential could have a zero 
at finite values of the fields. However, the F-term potential, as we showed 
above, vanishes only at infinity. Therefore, one expects that supersymmetry
is broken for generic values of the Fayet-Iliopoulos term; it is only possible 
to have  vanishing D- and F-term 
potentials for infinite (negative) value of the FI term. For finite values of the 
orbifold blow-up parameter, however, supersymmetry is always
broken.

We also note that if we  consider the branch of moduli space with
8 mod 6 D3 branes at the origin, i.e. the $SU(4)$ theory, we would find that
the theory has  restored supersymmetry for finite values of the blow-up
parameter (such that the one-loop Fayet-Iliopoulos term is cancelled).
This is because, in contrast to the $SU(5)$ model, the $SU(4)$ theory has
a branch of moduli space where no dynamical superpotential is 
generated---this can be inferred from
 \cite{IS} by noting that the $SU(4)$ theory 
with three ${\bf 6}$'s  is 
equivalent to the $SO(6)$ theory with three vectors. 
The breaking of 
supersymmetry is then purely D-term and vanishes for vanishing FI parameter.

We note that in the above discussion of supersymmetry breaking  we ignored 
the closed string modes. 
Once these are allowed to fluctuate, there are, as usual,
runaway directions along which supersymmetry is restored. We leave a detailed
investigation of this for  future work.

I  would like to thank Ken Intriligator and Witek Skiba for discussions.
This work was supported by DOE contract no. DOE-FG03-97ER40506. 
Additional support of the European Community TMR Programme and the 
National Science Foundation is  greatfully acknowledged.

\nc{\ib}[3]{ {\em ibid. }{\bf #1} (19#2) #3}
\nc{\np}[3]{ {\em Nucl.\ Phys. }{\bf #1} (19#2) #3}
\nc{\pl}[3]{ {\em Phys.\ Lett. }{\bf B#1} (19#2) #3}
\nc{\pr}[3]{ {\em Phys.\ Rev. }{\bf D#1} (19#2) #3}
\nc{\prep}[3]{ {\em Phys.\ Rep. }{\bf #1} (19#2) #3}
\nc{\prl}[3]{ {\em Phys.\ Rev.\ Lett. }{\bf #1} (19#2) #3}

\end{document}